\documentclass[preprint,showpacs,amsmath,amssymb]{revtex4}
\usepackage{graphicx}
\usepackage{dcolumn}
\usepackage{bm}

\begin{document}

\title{Dynamics of quantum adiabatic evolution algorithm for\\ Number Partitioning}

\author{Vadim N. Smelyanskiy\footnote{NASA Ames Research Center, MS 269-3, Moffett Field, CA 94035-1000}}
\email{vadim@email.arc.nasa.gov}
\author{Udo v. Toussaint\footnote{Max-Planck-Institute for Plasma Physics, Boltzmannstr, 2 D-85748 Garching}}
\author{Dogan A. Timucin$^{*}$}
\noaffiliation

\date{\today}

\begin{abstract}
   We have developed a general technique to study the dynamics of the
  quantum adiabatic evolution algorithm applied to random
  combinatorial optimization problems in the asymptotic limit of
  large problem size $n$. We use as an example the NP-complete
  Number Partitioning problem and map the algorithm dynamics to that
of an auxilary quantum spin glass system with the slowly varying Hamiltonian.
We use a Green function method to obtain the  adiabatic eigenstates and the
minimum excitation gap, $g_{\rm min}={\cal
O}(n\,2^{-n/2})$,
corresponding to the exponential complexity of the
  algorithm for Number Partitioning.  The key element of the analysis is the conditional
  energy distribution computed for the set of all spin
  configurations generated from a given (ancestor) configuration by simulteneous
  fipping of a fixed number of spins. For the problem in question
 this distribution is shown to depend on the ancestor spin
  configuration only via a certain parameter related to the energy
  of the configuration.  As the result, the algorithm dynamics can be described in
   terms of  one-dimenssional quantum diffusion in the
 energy space. This effect provides a general limitation on the power of
a quantum adiabatic computation in random optimization problems.
Analytical results are in  agreement with the numerical
  simulation of the algorithm.
\end{abstract}

\pacs{03.67.Lx,89.70.+c}

\maketitle

\section{\label{sec:Intro} Introduction}

 Since the discovery by Shor \cite{Shor} nearly a
decade ago of a quantum algorithm for efficient integer
factorization there has been a rapidly growing interest in the
development of new quantum algorithms capable of solving
computational problems that are practically intractable on
classical computers. Perhaps the most notable example is that of a combinatorial
optimization problem (COP). In the
simplest case the task in COP is to minimize the cost function
(``energy") $E_{\bf z}$ defined on a set of $2^n$ binary strings
${\bf z}=\{z_1,\ldots, z_n \}$ $z_j=0,1$, each containing $n$
bits. In quantum computation this cost function corresponds to a
Hamiltonian $H_P$
\begin{eqnarray}
&& H_P=\sum_{\bf z} E_{\bf z}  | {\bf z}\rangle   \langle {\bf z}|
\label{HP} \\ && |{\bf z}\rangle = | z_1\rangle_1\,
\otimes|z_2\rangle_2\,\otimes\cdots\otimes|z_n\rangle_n .
\nonumber
\end{eqnarray}
\noindent where $z_j=0,1$ and the summation is over
$2^n$ states $|{\bf z}\rangle$  forming the computational basis of a
quantum computer with $n$ qubits. State $|z_j\rangle_j$ of the
$j$-th qubit is an eigenstate of the Pauli matrix $\hat{\sigma}_z$
with eigenvalue $S_j=1-2z_j \,\,(S_j=\pm 1)$. It is clear from
the above that the ground state of $H_P$ encodes the solution to the
COP with cost function $E_{\bf z}$.

COPs have a direct analogy in physics, related to finding ground
states of classical spin glass models. In the example above bits
$z_j$ correspond to Ising spins $S_j$.  The connection between the
properties of frustrated disordered systems and the structure of
the solution space of complex COPs has been noted first by Fu and
Anderson \cite{Anderson}. It has been recognized \cite{Parizi}
that many of the spin glass models are in almost one-to-one
correspondence with a number of COPs from theoretical computer
science that form the so-called NP-complete class \cite{Garey}. This
class contains hundreds of the  most common computationally hard
problems encountered in practice, such as constraint satisfaction,
traveling salesmen, and integer programming. NP-complete
problems are characterized in the worst cases by exponential
scaling of the running time or memory requirements with the
problem size $n$. A special property of the class is that any
NP-complete problem can be converted into any other NP-complete
problem in polynomial time on a classical computer; therefore, it
is sufficient to find a deterministic algorithm that can be
guaranteed to solve all instances of just one of the NP-complete
problems within a polynomial time bound. It is widely
believed, however, that such an algorithm  does not exist on a classical
computer; whether it exists on a quantum computer is one of the
central open questions. Ultimately, one can expect that the
behavior of new quantum algorithms for COPs and their
complexity will be closely related to the properties of quantum
spin glasses.


Recently, Farhi and co-workers suggested a new  quantum algorithm
for solving combinatorial optimization problems  which is based on
the properties of quantum adiabatic evolution \cite{Farhi}.
Running of the algorithm for several NP-complete problems has been
simulated on a classical computer using a large number of randomly
generated problem instances that are believed to be
computationally hard for classical algorithms
\cite{FarhiSc,FarhiSat,FarhiCli,Hogg}. Results of these numerical
simulations for relatively small size of the problem  instances (
$n \leq$ 20) suggest a {\em quadratic} scaling law of the run time
of the quantum adiabatic algorithm with $n$.
Furthermore, it was shown in \cite{Vazirani02} that the previous query
complexity argument that led to the exponential lower bound for
unstructured search \cite{Bennett} cannot be used to rule out the
polynomial time solution of NP-complete Satisfiability problem by
the quantum adiabatic algorithm.

In \cite{Vazirani01,annealing,Vazirani02,farhimeas,BS} special
symmetric cases of COP were considered where symmetry of the problem
allowed the authors to describe the true asymptotic behavior
($n\rightarrow \infty$) of the algorithm.  In certain examples
considered in \cite{Farhi,annealing} the quantum adiabatic algorithm
finds the solution in time polynomial in $n$ while simulated annealing
requires exponential time.  This effect occurs due to the special
connectivity properties of the optimization problems that lead to the
relatively large matrix elements for the spin tunneling in transverse
magnetic field between different valleys during the quantum adiabatic
algorithm.  In the examples considered in \cite{annealing} the
tunneling matrix element scales polynomially with $n$. On the other
hand, in simulated annealing different valleys are
connected via classical activation processes for spins with
probabilities that scale exponentially with $n$. It was also shown for
certain simplified examples \cite{farhimeas,BS}, that quantum
adiabatic algorithm can be modified to completely suppress the
tunneling barriers even if the corresponding classical cost function
has  local minima well separated in the space of spin configurations.

However, so far there are no study on the true asymptotic behavior of
the algorithm for the general case of randomly generated hard
instances of NP-complete problems. Also there are no analysis of the
limitations of the quantum adiabatic computation arising from the
intrinsic properties of disorder and frustration in this problems.
Such analysis is of the central interest in this paper.

In Sec.~\ref{sec:NPP} we introduce the random Number Partitioning
problem and describes conditional cost distributions (neighborhood
properties) in this problem.  In Sec.~\ref{sec:QAEA} we describe the
concept of quantum adiabatic computation applied to combinatorial
optimization problems and introduce a Green function method for the
analysis of the minimum gap. In Sec.~\ref{sec:mingap} we describe the
effect of quantum diffusion in the algorithm dynamics, derive the
scaling for the minimum gap and the complexity of the algorithm for
the random Number Partitioning problem.  We also obtain the
scaling of the minimum gap numerically from the form of the
cumulative density of the adiabatic eigenvalues at the
avoided-crossing point.  In Sec.~\ref{sec:udo} we discuss the results of
the simulations of the time-dependent Schr{\" o}dinger equation to
simulate quantum adiabatic computation for Number Partitioning and
obtain its complexity numerically.

\section{\label{sec:NPP} Number Partitioning Problem}
Number Partitioning Problem (NPP)  is one of the six basic
NP-complete problems that are at the heart of the theory of
NP-completeness \cite{Garey}. It can be formulated as a
combinatorial optimization problem: $\quad$ Given a sequence of
positive numbers $\{a_1,\ldots, a_n\}$ find a partition, i.e.
two disjoint subsets ${\cal A}$ and ${\cal A}^{\prime}$, such that
the residue
\begin{equation}
E=\left | \sum_{a_j \in {\cal A}}a_j -\sum_{a_j \in {\cal
A}^{\prime}}a_j\right|\label{residue}
\end{equation}
\noindent
is minimized.
In NPP we search for the bit strings ${\bf z}=\{z_1,\ldots,
z_n\}$ (or corresponding Ising spin configurations ${\bf
S}=\{S_1,\ldots,S_n\}$)  that minimize the energy or cost function
$E_{\bf z}$
\begin{equation}
E_{\bf z}=\left | \Omega_{\bf S}\right |, \quad \Omega_{\bf
S}=\sum_{j=1}^{n}a_j S_j, \,\,S_j=1-2z_j, \label{omega}
\end{equation}
\noindent where $S_j=1$ ($z_j=0$) if $a_j\in {\cal A}$ and $S_j=-1$
($z_j=1$) if $a_j \in {\cal A}^{\prime}$.
The partition ${\bf S}$ with minimum residue
  can also be viewed as the ground state of the Ising spin glass,
 $-\Omega_{\bf S}^{2}$, corresponding to  the Mattis-like
antiferromagnetic coupling, $J_{ij}=-a_i\,a_j$.

NPP has many practical applications including multiprocessor
scheduling \cite{ms}, cryptography \cite{cr}, and others.
The best deterministic heuristical algorithm for NPP, the differencing
method of Karmakar and Karp \cite{Karp}, can find with high
probability solutions whose energies are of the order
$1/n^{\alpha\,\log\,n}$ for some $\alpha > 0$.  The interest in NPP
also stems from the remarkable failure of a standard simulated
annealing algorithm for the energy function (\ref{omega}) to find
good solutions, as compared with the solutions found by deterministic
heuristics \cite{johnson}.  The apparent reason for this failure is
due to the existence of order $2^n$ local minima whose energies are of
the order of $1/n$ \cite{Ferreira} which undermines the usual strategy
of exploring the space of the spin configurations ${\bf S}$ through
single spin flips.

The computational complexity of random instances of NPP depends on the
number of bits $b$ needed to encode the numbers $a_j$. In what follows
we will analyze NPP with independent, identically distributed (i.i.d.)
random b-bit numbers $a_j$. Numerical simulations show
\cite{Gent,Korf,MertensCompleteAnytime} that solution time grows
exponentially with $n$ for $n \ll b$ then decreases steeply for $n
\gtrsim b$ (phenomenon of \lq\lq peaking") and eventually grows
polynomially for $n \gg b$. The transition from the ``hard" to
computationally ``easy" phases at $n\approx b$ has features somewhat
similar to phase transitions in physical systems
\cite{Mertens98}. The detailed theory  of the phase transition in NPP
was given in Refs.~\cite{Borgs1,Borgs2}.
If one keeps the parameter $\xi=b/n$
fixed and lets $n \rightarrow \infty$ then instances of NPP
corresponding to $\xi >$ 1 will have no perfect partitions with high
probability. On the other hand for $\xi <$ 1 number of perfect
partitions will grow exponentially with $n$.  Transitions of this kind
were observed in various NP-complete problems \cite{AI96}. In what
follows we will focus on the computationally hard regime $\xi \gg 1$.

\subsection{\label{spp.den} Distribution of signed partition residues}

The  values of individual energies are
random and depend on the particular instance of NPP (i.e., the set
of numbers $a_j$). However on a  coarse-grained scale (i.e. after
averaging over individual energy separations) the form of the
typical energy distribution is described by some universal
function for randomly generated problem instances. 
We introduce for a given set of randomly sampled numbers $a_j$ a
coarse-grained  distribution function of signed partition residues
$\Omega_{\bf z}$ (\ref{omega})
\begin{equation}
 P(\Omega)= 2^{-n} \frac{1}{\Delta \Omega}\int_{\Omega-\Delta \Omega/2}^{\Omega+\Delta \Omega/2}
 d\eta \sum_{{\bf z}\in \{0,1\}^n }\delta(\eta-\Omega_{\bf z}).
 \label{rho}
\end{equation}
\noindent Here $\delta(x)$ is the Dirac delta-function; the sum is
over $2^n$ bit-strings ${\bf z}$ and $2^{-n}$ is a normalization
factor. In (\ref{rho}) we average  over an interval $\Delta \Omega$ of the
partition residues whose size is chosen
self-consistently, $\Delta \Omega \gg 2^{-n}/P(\Omega)$.  Using
(\ref{omega}) we can rewrite (\ref{rho}) in the form
\begin{eqnarray}
&&P(\Omega)=\frac{1}{2\pi}\int_{0}^{\infty}ds\,\zeta\left(\frac{\Delta
\Omega\, s}{2}\right)\,I(s)\,\cos (\Omega s ),\label{I}\\ &&
I(s)=\prod_{j=1}^{n}\cos (a_{j}\, s), \quad \zeta(x)=\sin(x)/x.
\nonumber
\end{eqnarray}
\noindent Here $\zeta(x)$ is a window function that imposes a cut-off
in the integral (\ref{I}) at $s \sim 2/\Delta \Omega$. For large
$n$ this integral can be evaluated using the steepest descent
method. In the following we shall assume that the $b$-bit numbers
$a_j$ are distributed inside of the unit interval $[0,1]$ and are
 integer multiples of $2^{-b}$, the smallest number that can be
represented with available number of bits $b$. We note that for
large $n$ the function $I(s)$ has sharp maxima (minima) with width
$\sim n^{-1/2}$ at the points $s_k=k\pi\,2^{b},\, k=0,1,\ldots;$
$|I(s_k)|=1$.  Only one saddle point at $s=0$ contributes to the
integral in (\ref{I}) due to coarse-graining of the distribution
(\ref{rho}). Indeed, it will be seen below that the window size
$2/\Delta \Omega$ can be chosen to obey the conditions ${\rm 1}
\ll n^{{\rm 1/2}}/\Delta \Omega \ll {\rm 2}^{n}$.  Therefore in
the case of high-precision numbers, $b\gg n$, saddle-points $s_k$
with $k>0$ lie far outside the window and their contributions can
be neglected (see also Appendix \ref{sec:subharmonics}). On the
other hand the window function $\zeta(x)$ can be replaced by unity
while computing the contribution from the saddle-point at $s=0$.
Finally we obtain for $|\Omega| \ll n$ (cf. \cite{Mertens2000})
\begin{eqnarray}
&& 
P(\Omega)=\frac{1}{\sqrt{2\,\pi\,\sigma^{2}(0) \,n}}\exp\left
(-\frac{\Omega^2}{2\sigma^{2}(0) \,n}\right)+{\cal O}(n^{-3/2})
\nonumber
\\ && \sigma^{2}(0)= \frac{1}{n}\sum_{j=1}^{n}a_{j}^{2} \qquad (E\ll
n)
.\label{PGauss}
\end{eqnarray} \noindent  The
coarse-grained distribution $P(\Omega)$ depends on the set of
$a_j$'s through a single self-averaging quantity $ \sigma(0)$ (cf.
\cite{Mertens98}).

One can also introduce
 the distribution $\tilde
P(E)$ of cost values (energies) $E_{\bf z}=|\Omega_{\bf z}|$.
 Due to the obvious
symmetry of the NPP, the cost function $E_{\bf z}$ in (\ref{omega})
does not change after flipping signs of all spins, $S_j\rightarrow
-S_j$. Therefore
\begin{equation}
\tilde{P}(E)=1/2 P(\pm E).\label{rhotilde} \end{equation}
\noindent We emphasize that, according to Eq.~(\ref{PGauss}) for a
typical set of high-precision numbers $a_j$ the energy spectrum in
NPP is quasi-continuous, and there are only two scales present in the
distribution $\tilde{P}(E)$: one is a \lq\lq microscopic" scale
given by the characteristic  separation of the individual
partition energies, $E_{\rm min}$, and another is given by the
mean partition energy $\langle E\rangle$ (or the distribution
width $\langle E^2\rangle^{1/2}$)
\begin{equation} E_{\rm min}\sim
\sigma(0)\, n^{1/2}\,\,2^{-n},\qquad \langle E^2 \rangle =
\frac{\pi}{2}\langle E\rangle^{2}= n\sigma^{2}(0). \label{meanE}
\end{equation}
\noindent This justifies the choice for $\Delta \Omega$ above that
corresponds to coarse-graining over many individual energy level
separations.

We note that the distribution $P(\Omega)$ (\ref{PGauss}) is
Gaussian for $E\ll n$ and can be understood in terms of a random
walk with coordinate $\Omega$ using Eq.~(\ref{omega}). The walk
begins at the origin, $\Omega=0$, and makes a total of $n$ steps. At
the $j$-th step $\Omega$ moves to the right or to the left by
\lq\lq distance" $2\,a_j$ if $S_j=1$ or $S_j=-1$, respectively. In
the asymptotic limit of large $n$ the result (\ref{PGauss})
corresponds to  equal probabilities of right and left moves  and
the distribution of step lengths coinciding with that of the set
of numbers $\{2\,a_j\}$.

Finally, the energy distribution function $P(E)$ of the form
(\ref{PGauss}),(\ref{rhotilde}) was previously obtained by Mertens
\cite{Mertens2000} using explicit averaging over the random
instances of NPP.   He also computed the partition function $Z(T)$
for a given instance of NPP at a small finite temperature $T$
using the steepest-descent method and summation over the
saddle-points $s_k=k\pi\,2^{b}$ similar to our discussion above
\cite{Mertens98} (in his analysis $k_B\,T$ played a role similar
to our regularization factor $\Delta \Omega$ in (\ref{rho}),(\ref{I})).

We emphasize however, that the approach in Ref.~\cite{Mertens98}
based on $Z(T)$ is necessarily restricted to the analysis of the \lq\lq
static" properties of NPP at $E \sim 2^{-n}$, i.e., the phase
transition in the number of perfect partitions \cite{Mertens98}
when the control parameter  $\xi=n/b$ crosses a critical value. On
the other hand distribution $P(\Omega)$ (\ref{rho}) introduces at
finite energies, as well as the conditional distribution
introduced in the next section also allow us to directly study the
intrinsic {\it dynamical} properties of the problem in question
such as the dynamics of its quantum optimization algorithms.

\subsection{Conditional distribution of signed partition residues}

Consider the set of bit-strings  ${\bf z'}$ obtained from a given
string ${\bf z}$ by flipping $r$ bits. The conditional
distribution of the partition residues $\Omega_{\bf z'}$
(\ref{omega}) in the $r$-neighborhood of ${\bf z}$  can be
characterized by its moments:
\begin{equation}
\langle \Omega^{k} \rangle = \binom{n}{r}^{-1}\sum_{{\bf z'} \in
\{0,1\}^n} \left(\Omega_{\bf z'}\right)^{k}\, \delta_{r,D({\bf
z'},{\bf z})}, \qquad k=1,2,\ldots \label{moments}
\end{equation}
\noindent Here  $\delta_{m,l}$ is a Kronecker delta and function
$D({\bf z},{\bf z^{\prime}})$ computes the number of bits that
take different values in the bit-strings ${\bf z}$ and ${\bf
z^{\prime}}$. It is the so-called Hamming distance between the
two strings
\begin{equation}
D({\bf z},{\bf
z^{\prime}})=\sum_{j=1}^{n}\left|z_j-z_{j}^{\prime}\right|.\label{hamming}
\end{equation}
\noindent The Hamming distance $r=D({\bf z},{\bf z^{\prime}})$
between the bit-strings is directly related to the overlap factor
$q$ between the corresponding spin configurations often used in
the theory of spin glasses \cite{Parizi,Mertens2000}:
\begin{equation}
q=\frac{1}{n}\sum_{j=1}^{n}S_{j}S_{j}^{\prime}=
1-\frac{2}{n}D({\bf z},{\bf z^{\prime}}).\label{overlap}
\end{equation}
\noindent (in what follows we shall use both quantities $r$ and
$q$). For $k=1,2$ in (\ref{moments}) one obtains after
straightforward calculation the first and second moments of the
conditional distribution
\begin{eqnarray}
&& \langle \Omega \rangle=q\,\Omega_{\bf z}, \label{m1}
\\ && \langle \Omega^{2}\rangle-\langle \Omega \rangle^2
=n\sigma^{2}(q)\,
\left(1+\frac{1}{n-1}\right)\,\left(1-\frac{1}{n}\,\frac{\Omega_{\bf
z}^{2}}{\langle E^2\rangle }\right), \label{m2}\\ &&
\sigma(q)=\sigma(0)\,(1-q^2)^{1/2}, \qquad q\equiv
1-\frac{2r}{n},\label{sigmaq}
\end{eqnarray}
\noindent where $\sigma(0)$ and $\langle E^2\rangle$ are given in
(\ref{PGauss}) and (\ref{meanE}), respectively.

The conditional distribution of $\Omega_{\bf z'}$  can also
be defined in a way similar to (\ref{rho})
\begin{equation}
P_{r,{\bf z}}(\Omega^{\prime})= \binom{n}{r}^{-1}\,\frac{1}{\Delta
\Omega'}\int_{\Omega'-\Delta \Omega'/2}^{\Omega'+\Delta \Omega'/2}
d\eta \sum_{\bf z^{\prime}\in\{0,1\}^n}\delta(\eta-\Omega_{\bf
z^{\prime}})\,\delta_{r, D({\bf z^{\prime}},{\bf z})}
\label{rhoOmegazr}
\end{equation}
\noindent where  averaging is over the small interval $\Delta
\Omega^{\prime}$ that, however, includes many individual values of
$\Omega_{z^{\prime}}$ for a given $r$.
It is clear from (\ref{m1}),(\ref{m2}) that the first two moments of
$P_{r,{\bf z}}(\Omega^{\prime})$ depend on ${\bf z}$ {\em only}
via the value of $\Omega_{\bf z}$. This does not hold true,
however, for the higher-order moments that depend on other
functions of ${\bf z}$ as well.  For example, $\langle
\Omega^3\rangle$ involves the quantity
$\sum_{j=1}^{n}a_{j}^{3}(1-2z_j)$, etc.

Our main observation is that in the asymptotic  limit of large $n$
the conditional distribution  $P_{r,{\bf z}}(\Omega^{\prime})$ is
well-described by the first two moments (\ref{m1}),(\ref{m2}).
Then, according to the discussion above, its dependence on ${\bf
z}$ is only via $\Omega_{\bf z}$. The detailed study of the higher
moments (\ref{moments}) will be done elsewhere. Here we use the
following intuitive approach relevant for analysis of the
computational complexity of the quantum adiabatic algorithm for
the NPP. We average $P_{r,{\bf z}}(\Omega^{\prime})$ over the
strings ${\bf z}$ with residues $\Omega_{\bf z}$ inside a small
interval $\Delta \Omega$ (containing, however, many levels
$\Omega_{\bf z}$). After such averaging the result,
$P_{r}(\Omega'|\Omega)$, can be written in the form
\begin{eqnarray}
&&P_{r}(\Omega'|\Omega) =\frac{P_{r}(\Omega',\Omega)}{P(\Omega)},
 \label{rhoOmegaOmega1}\\ &&
P_{r}(\Omega',\Omega)=2^{-n}\,\frac{1}{\Delta \Omega}
\int_{\Omega-\Delta \Omega/2}^{\Omega+\Delta \Omega/2}
 d\eta\,\sum_{{\bf z}\in\{0,1\}^n}\,\delta(\eta-\Omega_{\bf z})\,P_{r,{\bf z}}(\Omega')
,  \label{rhoOmegaOmega}
\end{eqnarray}
\noindent where $P(\Omega)$ is given in (\ref{PGauss}). We note
that Eq.~(\ref{rhoOmegaOmega1}) formally coincides with the
Bayesian rule expressing the conditional distribution function
$P_{r}(\Omega^{\prime}|\Omega)$ through the 2-point (joint)
distribution function $P_{r}(\Omega^{\prime},\Omega)$ and
the single-point distribution $P(\Omega)$ (\ref{PGauss}).

The explicit form of $P_{r}(\Omega|\Omega)$ is derived in Appendix
\ref{sec:conddist}  in a manner similar to the derivation of
$P(E)$ in Sec.~\ref{spp.den}. The results are presented in
Eqs.~(\ref{conditional}) and (\ref{theta}). They show that
$P_r(\Omega',\Omega)$ in the limit $n\gg 1$ is indeed well
described by its first two moments that corresponds precisely to
the expressions given in Eqs.~(\ref{m1}),(\ref{m2}) above. From
this we conclude that
\begin{equation}
P_{r,{\bf z}}(\Omega') = P_{r}(\Omega'|\Omega_{\bf
z}).\label{connection}
\end{equation}
\noindent

In the case $r=1$  there are $n$ strings ${\bf z'}$ at a Hamming
distance 1 from the string ${\bf z}$. Partition energies
corresponding to these strings equal $|\Omega_{\bf S}-2 a_j S_j|,
\,1\leq j\leq n$ (cf. (\ref{omega})). After the coarse-graining
over the energy scale ${\cal O}(1/n)$ in the range,
$|\Omega|,|\Omega'| \ll n$, the conditional distribution $P_{r,{\bf
z}}$ is a step function in the interval $\Omega_{\bf z}-\Omega'\in
\left[-2,2\right]$.  For $r=n-1$ one has  the same form of the
distribution  but for $\Omega_{\bf z}+\Omega'$. Both results
correspond to nearly equal distribution of spins between
between $\pm 1$ values. Then in the range of energies
$|\Omega^{\prime}|,|\Omega_{\bf z}| \lesssim 1$ one has:
\begin{equation}
P_{r,{\bf z}}(\Omega')\approx \bar{P}_r=  1/2 +{\cal
O}\left(\frac{1}{n}\right),\quad r=1,n-1\qquad (n \gg 1)
\label{rho1}
\end{equation}

\begin{figure}[bht]\hspace{-0.33in}
\includegraphics[width=3.7in]{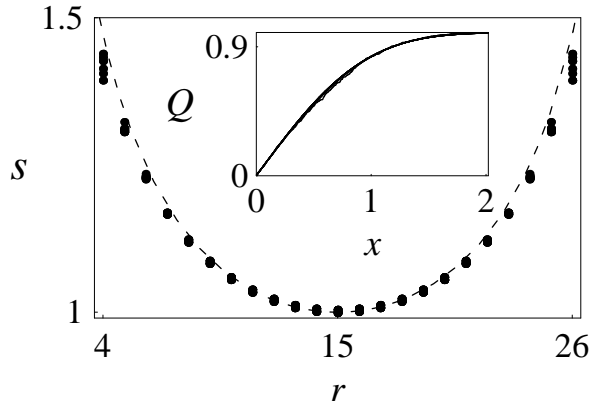}
\caption{\label{fig:density06} Plots of the (scaled) conditional
distribution (\ref{rhoOmegazr}) $s=\sigma(0) (2\pi
n)^{1/2}\,(\Delta\Omega)^{-1}\int_{0}^{\Delta\Omega}d\eta\,P_{r,{\bf
z}}(\eta)$  {\em vs} $r$ are shown with points. We use
coarse-graining window $\Delta\Omega$=0.3. Different plots
correspond to 29 randomly selected bit-strings ${\bf z}$ with
energies $|\Omega_{\bf z}| \in [0,0.3]$ for one randomly generated
instance of NPP with $n=30$  and $b=35$. For  $r, n-r\gg 1$ the
values of $s$ corresponding to different strings  are visually
indistinguishable from each other. Dashed line is a plot of
$\sigma(0)/\sigma(q)$ {\it vs} $\,\,r$ given in (\ref{m2})
($q=1-2r/n$).$\quad$   Insert: plots of
the integrated quantity given in (\ref{Qe}),
$Q=\frac{1}{2}\int_{0}^{\Omega}d\eta\,P_{r,{\bf z}}(\eta)$ {\em
vs} $x=\Omega/(\sigma(q)\sqrt{2n})$, for different values of
$r=2,\ldots,n/2$ and randomly selected bit-string ${\bf z}$ with
energy $|\Omega_{\bf z}|$ close to 0. All plots correspond to the
same instance of NPP as the main figure. Plots for different
values of $r$ are visually indistinguishable from each other and
from the theoretical curve given in (\ref{rhoEEa}).}
\end{figure}

For $r,n-r\gg 1$ distribution $P_{r,{\bf z}}(\Omega')$ has a
Gaussian form with a broad maximum  at $\Omega'=q\Omega_{\bf z}$
(cf. Eqs.~(\ref{m1}),(\ref{m2}),(\ref{conditional})) . Near the
maximum we have:
\begin{equation}
P_{r,{\bf z}}(\Omega')\approx \bar{P}_{r}=\frac{1}{\sqrt{2\pi
n\sigma^{2}(q)}},\qquad |\Omega'|,|\Omega_{\bf z}|\ll
n^{1/2}\sigma(q) . \label{nr}
\end{equation}
\noindent We studied the conditional distribution in NPP
numerically as well (see Fig.\ref{fig:density06} and
Sec.\ref{sec:numerics}). The results are in good agreement with
theory even for modest values of $n\leq 30$.

The characteristic spacing between the values of the partition residues
in the subset of strings ${\bf z^{\prime}}$ with $D({\bf
z^{\prime}},{\bf z})=r$ is $1/(\bar{P}_{r}\,\binom{n}{r})$ for not
too large $E_{\bf z},E_{\bf z^{\prime}}$ (see above). This spacing
decreases exponentially with the magnitude of the  string overlap
factor, $|q|=|(n-2r)/n|$.  The hierarchy of the subsets corresponding
to different values of $|q|$ form a specific structure of NPP. We
note that the distribution of partition residues within the
hierarchy is nearly independent of the ancestor string ${\bf z}$
in a broad range of energies $E'\lesssim n^{1/2}$  where
$P_{r,{\bf z}}(E')\approx \bar{P}_r$.  One can see that the
 magnitude of the overlap factor $q$ between two strings
with energies within a given interval  $[0,E]$ is limited by some
typical value $\bar{q}$ satisfying the following  equation:
\begin{equation}
E\binom{n}{r}\bar{P}_{r}= 1,\qquad
|\bar{q}|=1-2\frac{r}{n}.\label{sa}
\end{equation} The smaller $E$ is, the smaller $|\bar{q}|$ is: strings that
are close in energy are far away in the configuration space. This property
gives rise to an exponentially large number of local minima for
small values of $E_{\bf z}$ that are far apart in  the configuration space.
For example, strings with $E_{\bf z}\sim E_{\rm min}$ typically
correspond to $|q|={\cal O}(1/n)$, they can be obtained from each
other only by simultaneously flipping clusters with $\sim n/2$
spins.

Eq.~(\ref{sa}) describes the dynamics of a local search heuristic
(e.g., simulated annealing). It shows that the average cost value
$E$ during the search decreases  no faster than ${\cal O}(1/M)$
where $M={\cal O}\left(\binom{n}{r}\right)$ is the number  of
generated configurations. This result coincides with that obtained
in \cite{Mertens2000} using  a different approach. It says that
any classical local search heuristic for NPP cannot be faster than
random search. Indeed, during local search the information
about the \lq\lq current" string ${\bf z}$ with $E_{\bf z}\lesssim
1$  is being lost, on average, after one spin flip (cf.
Eqs.~(\ref{rho1}),(\ref{nr})). We show below that precisely this
property of NPP also leads  to the complexity of the quantum
adiabatic algorithm corresponding to that of a quantum random
search.

We note that one can trivially break the symmetry of NPP
mentioned above by introducing an extra number $a_0$ and placing
it, say, in the subset ${\cal A}$. In this case different
partition energies will still be encoded by  spin configurations
${\bf S}=\{S_1, \ldots, S_n\}$ (or corresponding bit-strings ${\bf
z}$) with $\Omega_{\bf S}=a_0+\sum_{j=1}^{n}S_j\,a_j$ and  $E_{\bf
z}=|\Omega_{\bf S}|$ (cf. \ref{omega}). We shall adopt this
approach in the analysis of the performance of the quantum
adiabatic algorithm for NPP given below.

\section{\label{sec:QAEA} Quantum Adiabatic Evolution Algorithm}

In the quantum adiabatic algorithm  \cite{Farhi} one specifies
the time-dependent Hamiltonian $H(t)=\tilde H(t/T)$
\begin{equation}
\tilde H(\tau) = (1-\tau)\, V+ \tau\, H_P, \label{H}
\end{equation}
\noindent where $\tau=t/T$ is dimensionless \lq\lq time". This
Hamiltonian guides the quantum evolution of the state vector
$|\psi(t)\rangle$  according to the Schr{\" o}dinger equation $
i\,{\partial |\psi(t)\rangle \partial t}  =H(t) |\psi(t)\rangle$
from $t=0$ to $t=T$, the {\em run time} of the algorithm (we let
$\hbar=1$). $H_P$ is the \lq\lq problem" Hamiltonian  given in
(\ref{HP}). $V$ is a \lq\lq driver" Hamiltonian, that is designed
to cause transitions between the eigenstates of $H_P$. In this
algorithm one prepares the initial state of the system $\psi(0)$
to be  the ground state of $\tilde H(0)=V$.
In the simplest case
\begin{equation}
 V = -\sum_{j=1}^{n} \sigma_{x}^{j},\quad
|\psi(0)\rangle =2^{-n/2}\sum_{\bf z}|{\bf z}\rangle, \label{V}
\end{equation}
\noindent  where $\sigma_{x}^{j}$ is a Pauli matrix for $j$-th
qubit.  Consider instantaneous eigenstates
$|\phi_\eta(\tau)\rangle$ of $\tilde H(\tau)$  with energies
$\lambda_\eta(\tau)$ arranged in nondecreasing order at any
value of $\tau\in(0,1)$
\begin{equation}
\tilde  H |\phi_{\eta}\rangle = \lambda_{\eta} |\phi_{\eta}\rangle,
\quad \eta=0,1,\ldots,2^{n}-1.\label{adiab}
\end{equation}\noindent
 Provided the value of $T$ is large enough and there is a finite gap
for all $t\in(0,T)$  between the
 ground and excited state energies,
 $g(\tau)=\lambda_1(\tau)-\lambda_0(\tau)>0$,
 quantum evolution is adiabatic and the state of the system
 $|\psi(t)\rangle$ stays
close to an instantaneous ground state, $|\phi_0(t/T)\rangle$ (up
to a phase factor). Because  $H(T)=H_P$ the final state
$|\psi(T)\rangle$ is close to the ground state
$|\phi_0(\tau=1)\rangle$ of the problem Hamiltonian. Therefore a
measurement performed on the quantum computer at $t=T\, (\tau=1)$
will find one of the solutions of COP with large probability.

There is a broad class of COPs from theoretical Computer Science
where the number of distinct values of a cost function scales
polynomially in the size of an input $n$.  An example is the
Satisfiability problem in which the cost $E_{\bf z}$ of a given
string ${\bf z}$ equals the number of constrains violated by
the string.  For those problems, the spectrum of $H(\tau)$, at the
beginning ($\tau\approx 0$) and at the end ($\tau\approx 1$) of
the algorithm, consists of a polynomial number of well-separated
energy levels. Quantum transitions away from the adiabatic ground
state occur most likely near the avoided-crossing points
$\tau\approx \tau^{*}$ where the energy gap $g(\tau)$ reaches its
minima \cite{Hogg}. Near the avoided-crossing points, the spectrum of
$H(\tau)$ is quasi-continuous, with the separation between
individuals eigenvalues scaled down with $n$.  The probability of
a quantum transition, $1-|\langle
\psi(t)|\phi_0(t/T)\rangle|^{2}_{t=T}$, is small provided that
\begin{equation}
T\gg  \frac{ |\langle \phi_{1}|\tilde
H_{\tau}|\phi_{0}\rangle|_{\tau=\tau^{*}} }{g_{\rm min}^{2}}
,\quad g_{\rm min}=\min_{0\leq \tau\leq
1}\left[\lambda_1(\tau)-\lambda_0(\tau)\right],
\label{mingap}
\end{equation}
\noindent ($\tilde{H}_{\tau} \equiv d\tilde{H}/d\tau$). The
fraction in (\ref{mingap}) gives an estimate for the required
runtime of the algorithm and the task is to find its asymptotic
behavior in the limit of large $n \gg 1$.  The numerator in
(\ref{mingap}) is less than the largest eigenvalue of
$\tilde{H}_{\tau}=H_P - V$, typically polynomial in $n$
\cite{Farhi}. However, $g_{\rm min}$ can scale down exponentially
with $n$ and in such cases the runtime of the quantum adiabatic
algorithm will grow exponentially with the size of COP.

\subsection{\label{sec:implementation} Implementation of QAA for NPP}

As suggested in \cite{Farhi} the quantum adiabatic algorithm  can
be recast within the conventional quantum computing paradigm using
the technique introduced by Lloyd \cite{Lloyd}. Continuous-time
quantum evolution  can be approximated by a time-ordered product
of unitary operators, $e^{-i\,(1-\tau_k)V\delta}\,e^{-i \tau_k H_P
\delta}$, corresponding  to  small time intervals
$(t_k,t_k+\delta)$. Operator $e^{-i\,(1-\tau_k)V\delta}$ typically
corresponds to a sequence of 1- or 2-qubit gates (cf. (\ref{V})).
Operator $e^{-i \tau_k H_P\delta}$ is diagonal in the
computational basis $|{\bf z}\rangle$ and corresponds to phase
rotations by angles $E_{\bf z}\delta$. Since in the case $n \ll
b$, the average separation between the neighboring values of
$E_{\bf z}$ is $1/P(E)={\cal O}(2^{-n})$, the quantum device
would need to support a very high precision in its physical parameters
(like external fields, etc.) to control small ${\cal O}(2^{-n})$ differences in
phases. Since this precision scales with $n$
exponentially it would strongly restrict the size of an instance
of NPP that could be solved on such a quantum computer. This
technical restriction is generic for COPs that involve a
quasi-continuous spectrum of cost-function values. Among the other examples
are many Ising spin glass models in physics
(e.g., the Sherrington-Kirkpatrick model \cite{Parizi}).
To avoid this restriction we introduce a new oracle-type cost function
${\cal E}_{\bf z}$ that returns a set of values
\begin{equation}
{\cal E}_{\bf z}=c(\Omega_{\bf z}), \qquad c(x)\longrightarrow \{\varepsilon_{0}, \varepsilon_{1}, \ldots, \varepsilon_{M}\} \qquad (\varepsilon_{k+1}>\varepsilon_{k}), \label{ek}
\end{equation}
\noindent
that can be stored using a relatively small number of bits ${\cal
  O}(\log n)$. For example, we can divide an interval of partition
energies $(0, B)$, $B=\sum_{j=0}^{j=n}a_{j}$ into bins whose sizes grow
exponentially with the energy. Then the  new cost will take one value per bin
\begin{eqnarray}
&& c(x) =\varepsilon_k\equiv-M+k \quad\qquad\,{\rm for}\,\,
\omega_{k}\leq |x| < \omega_{k+1},\nonumber \\ &&
\omega_{k}=(2^k-1)\,\Delta,\quad k=0,\ldots,\, M. \label{omegaK}
\end{eqnarray}
\noindent The last bin is $\omega_M \leq |\Omega_{\bf z}|\leq B$ where we
have ${\cal E}_{\bf z}=\varepsilon_M=0$. The value of the cutoff $\omega_{M}\leq B$
is discussed below.
In this example the Hilbert space of $2^n$ states $|{\bf z}\rangle$ is divided
into $M+1$ subspaces ${\cal L}_{k}$, each determined by
Eq.~(\ref{omegaK}) for a given $k$
\vspace{-0.07in}
\begin{equation}
H_P=\sum_{k=0}^{M}\varepsilon_{k}\sum_{{\bf z}\in{\cal L}_k}|{\bf
z}\rangle\langle{\bf z}|.\label{HPnew}
\end{equation}
\noindent Note that subspace ${\cal L}_0$ contains the solution(s)
to NPP. Dimension $d_0$ of ${\cal L}_0$  is controlled by the
value of $\Delta$ in (\ref{omegaK}) which is another control
parameter of the algorithm. We set $\Delta=2^{-n}\,K/P(0)$ where the
integer $K\approx d_0\gg 1$ is independent of $n$ and determines
how many times on average one needs to repeat the quantum
algorithm in order to obtain the solution to NPP  with probability
close to 1.

Operator $H_P$ projects any state $|\psi\rangle$ onto the states
with partition residues in the range $0\leq |\Omega_{\bf z}| < \omega_M$.
If we choose
\begin{equation} 1 \alt \omega_{M}\ll \langle E\rangle,\label{criterion_ld}
\end{equation} \noindent then the distribution  function
(\ref{PGauss}) is nearly uniform for $|\Omega_{\bf z}|\leq \omega_M$.
Therefore the dimensions of the subspaces ${\cal L}_k$ grow
exponentially with $k$: $d_k=d_0\, 2^k$ for $k<M$. This
simplification would not affect the complexity of a quantum
algorithm that spends most of its time in \lq\lq annealing"  the
system to much smaller partition residues, $\omega_M \gg |\Omega_{\bf
z}|\sim E_{\rm min}={\cal O}(n^{1/2}\,2^{-n})$.

We note that the new discrete-valued cost function defined in
(\ref{omegaK}) is  non-local. Unlike problems such as Satisfiability,
it cannot be represented by a sum of terms each involving a small
number of bits. To implement a unitary operator $e^{-i \tau_k
H_P\delta}$ with $H_P$ given in (\ref{HPnew}) one needs to
implement the following classical function on a quantum computer
\begin{equation}
{\cal E}_{\bf z}=\Theta(\omega_{M}-|\Omega_{\bf
z}|)\,\left[\log_{2}\left(\frac{\Delta+|\Omega_{\bf
z}|}{\Delta+\omega_{M}}\right)\right], \qquad \Omega_{\bf
z}=\sum_{j=1}^{n}a_j(1-2z_j). \label{Epsilon}
\end{equation}
\noindent Here $[x]$ denotes the integer part of a number $x$;
$\Theta(x)$ is the theta-function ($\Theta(x)=1$ for $x\geq 0$ and
$\Theta(x)=0$ for $x<0$). The implementation of (\ref{Epsilon})
with quantum circuits involves, among other things,  the addition
of $n$ numbers  together with their signs to compute $\Omega_{\bf
z}$, and taking the discrete logarithm of a $b$-bit number with
respect to  base $2$. These operations can be performed using a
number of quantum gates that is only polynomial in $n$ and $b$
(cf. \cite{Shor} for the implementation of the discrete
logarithm).

Since the implementation of a cost function
(\ref{ek}),(\ref{Epsilon}) does not add an exponential overhead to
the complexity of QAA the feasibility of this algorithm for NPP
depends on the scaling of the minimum gap $g_{\min}$ with $n$.

\subsection{\label{sec:stationary} Stationary Schr{\" o}dinger equation for adiabatic eigenstates}

We now solve the stationary Schr{\" o}dinger equation (\ref{adiab})
and obtain the minimum gap  $g_{\rm min}$ (\ref{mingap}) in the
asymptotic limit $n\rightarrow \infty$. To proceed we need to
introduce a new basis of states $|{\bf x}\rangle= |
x_1\rangle_{1}\,
\otimes|x_2\rangle_{2}\,\otimes\cdots\otimes|x_n\rangle_{n}$ where
state $|x_j\rangle_{j}$  is an eigenstate of the Pauli matrix
$\hat{\sigma}_{x}$ for the $j$-th qubit with  eigenvalue $1-2x_j
=\pm 1$. Driver Hamiltonian $V$ can be written in the following
form:
\begin{equation}
V=\sum_{m=0}^{n}\,V_{m}\,{\cal I}^{m},\,\,\,\, {\cal
I}^{m}=\sum_{x_1+\cdots+x_n=m} |{\bf x}\rangle\langle{\bf x}|.
\label{Dm}
\end{equation}
For a particular case given in Eq.~(\ref{V}) we have $V_{m}=2m-n$.
Matrix elements of ${\cal I}^{m}$ in a basis of states $|{\bf
z}\rangle$ depend only on the Hamming distance $D( {\bf z},{\bf
z^{\prime}})$  between the strings ${\bf z}$ and ${\bf
z^{\prime}}$
\begin{equation}
\langle {\bf z}| {\cal I}^{m}|{\bf z^{\prime}}\rangle =
 I^{m}_{D({\bf z},{\bf z^{\prime}})}, \label{Dzz}
\end{equation}
\vspace{-0.2in}
\begin{equation}
I^{m}_{r}=2^{-n}\sum_{q=0}^{n-r}\sum_{p=0}^{r}
\binom{n-r}{q}\binom{r}{p}(-1)^{p}\,\Delta_{m,\,q+p}.\label{Irm}
\end{equation}
\noindent We now rewrite  Eq.~(\ref{adiab}) in the form
\begin{equation}
|\phi\rangle = \frac{\tau}{\lambda-\alpha
V}\,H_P\,|\phi\rangle,\quad \alpha \equiv
\alpha(\tau)=1-\tau,\label{stSch}
\end{equation}
(we  drop the subscript $\eta$ indicating the number of a quantum
state and also the argument $\tau$ in $\phi$ and $\lambda$). From
(\ref{omegaK})-(\ref{stSch}) we obtain the equation for the
amplitudes $\phi_{\bf z}=\langle{\bf z}|\phi \rangle$  in terms of
the coefficients $I^{m}_{r}$
\begin{equation}
\left[1-\tau G_{0}\,c(\Omega_{\bf z})\right]\,\phi_{\bf z} = \frac{ \tau \Phi\,
2^{-n}}{\lambda-\alpha V_0}+\tau \sum_{{\bf z^{\prime}} \neq {\bf
z}}G_{D({\bf z},{\bf z^{\prime}})}\, \phi_{\bf z^{\prime}}\,
c(\Omega_{\bf z}), \label{Psi}
\end{equation}
\[\Phi=\sum_{\bf z^{\prime}}
c(\Omega_{\bf z^{\prime}})\, \phi_{\bf z^{\prime}},\nonumber
\]
\vspace{-0.2in}
\[
G_r\equiv G_{r}(\lambda)
=\sum_{m=1}^{n}\frac{I^{m}_{r}}{\lambda-\alpha V_{m}},\quad 0\leq
r\leq n.\nonumber
\]
\noindent Here we separated out a \lq\lq symmetric" term $\propto
2^{-n}\Phi$  corresponding to the coupling between the states
$|{\bf z}\rangle$ via the projection operator ${\cal I}^{0}$
(\ref{Dm}).

\section{\label{sec:mingap} Minimum gap analysis}

\subsection{\label{sec:cg} Coarse-graining of the transition matrix}

  We now make a key observation that $\phi_{\bf z}$ in
(\ref{Psi}) can be determined based on the properties of the
conditional distribution
 $P_{r,{\bf z}}(E)$  (\ref{rhoOmegazr}) and the form of the  Green function
$G_r(\lambda)$.
We sum  the Green function $G_{D({\bf z},{\bf z^{\prime}})}$ over
all possible transitions from a given  state ${\bf z^{\prime}}$ to states
${\bf z^{\prime}}\neq {\bf z}$  with energy $\varepsilon_k$. For not too large
partition residues of  the initial and final states we obtain
\begin{equation}
\sum_{ {\bf z}\in {\cal L}_k,\, {\bf z}\neq {\bf z^{\prime}}}
G_{D({\bf z},{\bf z^{\prime}})}(\lambda)\approx F_{k}(\lambda) +
f_{{\bf z^{\prime}},k}(\lambda)\label{F}
\end{equation}
\noindent 
 \vspace{-0.05in}
\begin{equation}
F_{k}(\lambda)=\frac{\mu\,s(\lambda)}{2^{M-k}},\,\,\,
  s(\lambda)=\int_{0}^{n}dr\,
\frac{\sigma(0)}{\sigma\left(1-2r/n\right)}\binom{n}{r} G_{r}(\lambda)\hspace{-0.01in}\label{Fk}
\end{equation}

\begin{equation}
|\Omega_{\bf z'}|,\,|\Omega_{\bf z}|\ll  \langle E\rangle, \qquad
 \mu=\frac{2\omega_M}{\pi\langle E\rangle}.\label{mu}
\end{equation}
\noindent
Function
$\sigma(q)$ above is defined in (\ref{sigmaq}) and $f_{{\bf
z^{\prime}},k}(\lambda)$ is a small correction described below. In
function $s(\lambda)$ we replaced summation over the integer
values of $r$ by an integral. It can be evaluated using the
explicit form of $G_r(\lambda)$ that decays rapidly with $r$. In
what follows we will be interested in the region $|\lambda-\alpha
V_0| \ll 1$ where
\begin{equation}
-2 \alpha \,G_{r}(\lambda)=\binom{n}{r}^{-1}
\sum_{m=1}^{n-r}\frac{2^{-n}\binom{n}{m+r}}{m}-2^{-n}\left(\ln
r+\gamma\right).\label{Jr}
\end{equation}
\noindent ($\gamma$ is Euler's constant) and $s(\lambda)\approx
-\ln 2/(2\alpha)$. We note that
\begin{equation} -2\alpha\, G_r(\lambda) \approx
-\left[(n/2-r)\binom{n}{r}\right]^{-1}, \quad n/2-r \gg 1.
\end{equation}\noindent
 Therefore the integrand in $s(\lambda)$ is
a smooth function of $r$ for $r\alt n/2$ and quickly decays to
zero for $r\agt n/2$. The contribution to the integral in
$s(\lambda)$ from the range of $r \ll n$ is small (${\cal
O}((r/n)^{1/2}$).

We note that term $F_{k}$ in (\ref{F}) provides an \lq\lq
entropic'' contribution to the sum in (\ref{F}). It comes from the
large number of states ${\bf z}\in {\cal L}_{k}$ corresponding to
large Hamming distances $r$ from the state ${\bf z^{\prime}}$, $ 1 \ll r
\alt n/2$.  Each state contributes a small weight, $G_{r}\propto
\binom{n}{r}^{-1}$, and number of states for a given $r$ is large,
 $(\omega_{k+1}-\omega_{k})\,\binom{n}{r}\,\bar{P}_r \gg 1$. Here $(\omega_{k+1}-\omega_{k})$
is an energy bin for the subspace  ${\cal L}_k$ and $\bar{P}_r$ is the
 conditional density of
states described in Sec. \ref{sec:NPP}.
The size of the bin  scales down
exponentially with $k$ (cf. (\ref{omegaK})) and so does the
entropic term $F_{k}$.  Below a certain cross-over value of $k$
one has  $|F_k| \ll |f_{{\bf z^{\prime}},k}(\lambda)|$. In this case the dominant
contribution to the sum (\ref{F}) comes from the states ${\bf
z}$ with small $r=D({\bf z},{\bf z^{\prime}})\sim 1$.  In particular for
 $k=0$ one can obtain
\begin{equation}
f_{{\bf z^{\prime} },0}(\lambda)
  \approx G_{1}(\lambda) \, \sum_{{\bf w}\in {\cal L}_0} \delta_{1,D({\bf z^{\prime}},{\bf w})}+
  {\cal O}( n^{-3}), \label{fzk}
\end{equation}
\noindent
where the higher-order terms correspond to $D({\bf z^{\prime}},{\bf
w})\geq$ 2. According to (\ref{Jr}),  $|G_{1}(\lambda)|\sim n^{-2}$ and
therefore $|f_{{\bf z},0}|$ is exponentially larger than the entropic term, $|F_0|\sim
\omega_0\sim d_0\,2^{-n}$. We note that, unlike the entropic term,
$f_{ {\bf z^{\prime}},0}$ strongly depends on ${\bf z^{\prime}}$  due to the
discreteness of the partition energy spectrum
($\omega_0\,n  \ll 1$). E.g., depending on a state ${\bf
z^{\prime}}$, in this case there could be either one or {\em none}
of the states ${\bf w }\in {\cal L}_{0}$ in the sum (\ref{fzk}) satisfying $D({\bf
z^{\prime}},{\bf w})= 1$.

\subsection{\label{sec:el} Extended and  localized eigenstates}

Based on the  discussion above we look for
solution of Eq.~(\ref{Psi}) in the following form:
\begin{equation}
\phi_{\bf z} = v(\Omega_{\bf z}) +u_{\bf z},   \qquad {\bf z}\notin {\cal L}_{0}, \label{ansatz}
\end{equation}
where we have explicitly separated out a part of the wavefunction $v(\Omega_{\bf z})$
 that depends on
${\bf z}$ only via the corresponding  value of the partition residue.
It  satisfies the following equations:
\begin{equation}
\left[1-\tau\,G_{0}(\lambda)\,c(\Omega)\right] \,v(\Omega) =
\frac{\tau \Phi\,2^{-n}}{\lambda-\alpha V_0} +\tau
\int_{\infty}^{\infty} d\Omega^{\prime}\,
v(\Omega')\,c(\Omega')\,\chi(\Omega',\Omega,\lambda), \label{v}
\end{equation}
\noindent \begin{equation}
\chi(\Omega',\Omega,\lambda)=\sum_{r=1}^{n}
\binom{n}{r}\,G_{r}(\lambda)\, P_{r}(\Omega'|\Omega).\label{chi}
\end{equation}
where $\Phi$ is  given in (\ref{Psi}) and function $c(x)$  takes a set
of discrete values (\ref{ek}).
Using  (\ref{Psi}),(\ref{ansatz}) and (\ref{v}) we obtain equations for $u_{\bf z}$
\begin{equation}
\left[1-\tau\,G_{0}(\lambda)\,\varepsilon_{k} \right] \, u_{\bf z} = \tau\,\sum_{k'=1}^{M}\varepsilon_{k'}\sum_{{\bf z^{\prime}}\in {\cal L}_k} G_{D({\bf z},{\bf z^{\prime}})}(\lambda)\,
u_{\bf z^{\prime}} + \tau \varepsilon_{0}\sum_{{\bf w}\in
{\cal L}_0} G_{D({\bf z},{\bf w})}(\lambda) \phi_{\bf w}, \qquad {\bf z}\in {\cal L}_k.\label{tilde}
\end{equation}
\noindent
Decomposition (\ref{ansatz}) is only applied to
amplitudes $\phi_{\bf z}$
with ${\bf z} \notin {\cal L}_0$.  The system of equations for the
components $v(\Omega)$ and $u_{\bf z}$ is closed by adding Eq.~(\ref{Psi})
for the amplitudes $\phi_{\bf w}$ with ${\bf w}\in {\cal L}_0$ (ground
states of the final Hamiltonian $H_P$) and taking (\ref{ansatz}) into
account.
We note that Eq.(\ref{v}) for  $v(\Omega)$ is coupled
to the rest of the equations only via the symmetric
term $\Phi$
\begin{eqnarray}
&& \Phi=\overline{\Phi}+\widetilde{\Phi}+\Phi_0\label{Phisym} \\
&&\overline{\Phi}=2^{n}\,\int_{-\infty}^{\infty}dx\, P(x)\, v(x)\,c(x),\label{phibar}\\
&&\widetilde{\Phi}=\sum_{k=1}^{M}\varepsilon_{k} \sum_{{\bf z}\in {\cal L}_k}
u_{\bf z}, \quad\Phi_0  =\varepsilon_0 \sum_{{\bf w}\in {\cal
L}_{0}}\phi_{\bf w}, \nonumber
\end{eqnarray}
\noindent
where distribution $P(\Omega)$ is given in (\ref{PGauss}).

\subsubsection{\label{sec:ld} Minimum gap estimate for $\omega_M\ll \langle E\rangle$}

We will analyze the above system of equations (\ref{ansatz})-(\ref{phibar}) assuming
that the cutoff frequency $\omega_M$ satisfies Eq.(\ref{criterion_ld}). This condition corresponds to the linear
region in the plot of the cumulative density of states given in insert to the Fig.~\ref{spp.den}.
According to Eqs.~(\ref{PGauss}),(\ref{rho1}), in this range the distribution functions $P(\Omega)\approx \bar{P}_{n/2}$
and $P_{r}(\Omega'|\Omega)\approx \bar P_r$ take nearly constant values and spectral function $\chi(\Omega',\Omega,\lambda)$
equals
\begin{equation}
\chi(\Omega',\Omega, \lambda) \approx \frac{s(\lambda)}{\sqrt{2\pi \,n\,\sigma^{2}(0)}}.\label{chi1}
\end{equation}
\noindent
where $s(\lambda)$ is given in (\ref{Fk}).
In this approximation, we can compute $\widetilde{\Phi}$ using equations for
$u_{\bf z}$ in (\ref{tilde}) and also the relations in
(\ref{F}), (\ref{Fk})
\begin{equation}
\widetilde{\Phi} = -\kappa\left(\tau\mu s(\lambda)\right
)\Phi_{0},\qquad \kappa(x)=\frac{x}{1+x}. \label{Phitilde}
\end{equation}
\noindent  In the initial stage of the
algorithm  the amplitudes $\phi_{\bf w}$ of the \lq\lq solution"
states are small $|\Phi_{0}|={\cal O}(2^{-n/2})$. According to (\ref{Phitilde}),
we also have  $|\widetilde{\Phi}|={\cal O}(2^{-n/2})$. Neglecting these terms and setting  $\Phi\approx
\overline{\Phi}$, Eq.~(\ref{v}) gives a closed-form algebraic
equation for $\lambda$
\begin{equation}
1+2\tau \mu\left(\frac{1}{\lambda-\alpha \,V_0}+s(\lambda)\right)=0.\label{transc}
\end{equation}
\noindent
Expanding  in a small parameter  $\mu \ll 1$
(cf.(\ref{criterion_ld}),(\ref{mu})), we obtain the eigenvalue
\begin{equation}
\lambda_{0}^{ i
}(\tau)\approx  \alpha(\tau)V_0-2\tau\mu -\frac{2(\tau\,\mu)^2\,\ln 2}{\alpha}  + {\cal O}(\mu^3)\qquad (\alpha \gg \mu),\label{li}
\end{equation}
\noindent
that accurately tracks the adiabatic ground state energy,
$\lambda_{0}(\tau)$, from $\tau=0$, up until small vicinity of the
avoided-crossing, $\tau\approx \tau^*$ (see below) where
$|\Phi_{0}|\sim 1$.

In the avoided-crossing region, branch $\lambda_{0}^{ i }(\tau)$
intersects with another branch,  $\lambda_{0}^{ f }(\tau)$, that
tracks $\lambda_{0}(\tau)$ in the interval of time following the
avoided-crossing, $\tau^{*} < \tau \leq 1$. This branch corresponds to
$\overline{\Phi} \ll \Phi_{0},\widetilde{\Phi}$. It can be
obtained from simultaneous solution of equations for
$u_{\bf z}$ (\ref{tilde}) and $\phi_{\bf w}$  that are
approximately decoupled from Eq.~(\ref{v}) after $\overline{\Phi}$ is
neglected. Keeping this term in (\ref{tilde}) gives rise to repulsion between
branches $\lambda^{i,\,f}_{0}(\tau)$ at $\tau=\tau^{*}$ that
determines the minimum gap $g_{\rm min}$ (see below).

To proceed, we obtain the equation for $\Phi_0$ by adding equations
for amplitudes $\phi_{\bf w}$ that correspond to different states
${\bf w}\in {\cal L}_0$ and neglecting the coupling between these
states  separated by large Hamming distances, $D({\bf w},{\bf
w^{\prime}})\sim n/2$. It can be shown using  Eqs.~(\ref{Psi})
and (\ref{fzk})-(\ref{tilde}) that $u_{\bf z}$ enters
equation for $\Phi_0$  through the term
\begin{equation} \tau^2\varepsilon_{0}\,
\sum_{{\bf z} \notin {\cal L}_0}{\cal E}_{\bf z}f_{{\bf
z},0}(\lambda) u_{\bf z},\label{self-energy}
\end{equation}
\noindent
which is is a self-energy term
corresponding to elementary bit-flip processes with initial and
final states belonging to the subspace ${\cal L}_0$ (loop
diagrams).

To express  $ u_{\bf z}$  in (\ref{self-energy}) through
$\phi_{\bf w}$ we solve Eq.~(\ref{tilde}) using order-by-order
expansion in a small parameter $n^{-1}$ (cf.  Eqs.~(\ref{F})-(\ref{fzk})
and discussion there).
In particular, one can show that to the leading order in
$n^{-1}$ the self-energy term  (\ref{self-energy}) is determined by lowest-order loops
with two bit flips that begin and end at ${\cal L}_0$.
Then after some transformations, the equation for
$\Phi_{0}$ takes the form
\begin{equation}
\Phi_{0}\left(\lambda-\tau\varepsilon_{0}-\frac{\tau\alpha^{2}\varepsilon_{0}}{\lambda}\sum_{{\bf
z^{\prime}}\notin {\cal L}_0} \frac{\delta_{1,D({\bf
z^{\prime}},{\bf w})}}{\lambda-\tau{\cal E}_{\bf
z^{\prime}}}\right) = \lambda\varepsilon_{0}\tau
d_{0}2^{-n} \left(\frac{\Phi}{\lambda-\alpha
V_0}+\overline{\Phi}\,s(\lambda)\right).\label{Phi0}
\end{equation}
\noindent Here $\alpha=1-\tau$ (cf. (\ref{stSch}) and $\bar{\Phi}$
is defined above. We now solve Eq.~(\ref{Phi0}) jointly with
(\ref{v}) and obtain a closed-form equation for $\lambda$. We
give it below in the region of interest $|\tau-1/2|\ll 1$
\begin{eqnarray}
&&\left(\lambda-\lambda^{\rm
i}_{0}(\tau)\right)\left(\lambda-\lambda^{\rm f}_0(\tau)\right)
=-n^2 2^{-n}\Delta^2/4 \label{switching} \\ &&\Delta \approx
d_{0}^{1/2} \left(1+\mu \tau^* \ln 2 +{\cal
O}(\mu^2)\right), \nonumber
\end{eqnarray}
\noindent  where the branch $\lambda^{\rm i}_{0}(\tau)$ is given above and
the branch $\lambda^{\rm f}_{0}(\tau)$ satisfies Eq.~(\ref{Phi0}) with r.h.s. there
set to zero,
\begin{equation}
\lambda^{\rm f}_{0}(\tau)\approx \tau \varepsilon_{0}-1/2,\quad |\tau-1/2|\ll 1.\label{lf}
\end{equation}
\noindent
Avoided-crossing in  (\ref{switching}) takes place at $\tau=\tau^*$
\begin{equation}
\quad\lambda^{\rm i}_0(\tau^*)=\lambda^{\rm
f}_0(\tau^*),\quad \tau^* \approx \frac{1}{2} + \frac{1}{4n} \log_{2}\frac{d_0}{\mu}.
\end{equation}
\noindent
The value of minimum gap between the two roots of (\ref{switching}) equals
\begin{equation}
g_{\rm
min}=n\,\Delta \,2^{-n/2}.\label{gmin}
\end{equation}
\noindent
where $\Delta$ is defined in (\ref{omegaK}).

Based on the
above analysis one can also estimate the matrix element
$|\langle \phi_{1}|\tilde
H_{\tau}|\phi_{0}\rangle|_{\tau=\tau^{*}}\sim n$.
Then from Eq.~(\ref{mingap}) (see  also discussion after Eq.~(\ref{HPnew}))
one can estimate the run-time   of the quantum adiabatic
algorithm
\begin{equation}
T\gg \frac{d_{0}\,|H_{\tau 01}^{*}|}{g_{\rm min}^{2}}={\cal O}((n\,d_{0})^{-1} 2^{n}).\label{lower_bound}
\end{equation}

It follows from the  above that eigenvalue
branch $\lambda^{\rm i}_{0}(\tau)$  corresponds to a  state,
\[ |\phi_{0}\rangle \approx \sum_{{\bf z}\in\{0,1\}^n} v(\Omega_{z}) |{\bf z}\rangle,\]
\noindent
which is {\em extended} in the space of the bit configurations $|{\bf z}\rangle$: according to
(\ref{v}) it contains
a large number (${\cal O}(2^n)$) of exponentially
small (${\cal O}(2^{-n/2})$) individual amplitudes.  This state originates at $\tau=0$ from the totally symmetric
initial state $|\psi(0)\rangle$ (\ref{V}). In  the  small region  $|\tau-\tau^*| \sim
g_{\rm min}$  it is transformed into the  state that corresponds to the eigenvalue branch
 $\lambda^{\rm f}_{0}(\tau)$ and is
 {\em localized} in Hamming distances $D({\bf z},{\bf w})$ near the subspace ${\bf w}\in {\cal L}_0$ containing the solution to NPP
$|\phi_{0}\rangle \approx \sum_{{\bf w}\in{\cal L}_0}|{\bf
w}\rangle$.
Minimum gap at the avoided-crossing is determined by the overlap
between the extended and localized states.

At later times $\tau > \tau^{*}$ a similar picture applies to the
avoided crossing of the extended-state energy $\lambda^{\rm
i}_{0}(\tau)$ with energies of localized states $\lambda^{\rm
f}_{k}(\tau)$ corresponding to ${\bf z}\in{\cal L}_k$ with $1\leq
k\ll n$ (excited levels of the final Hamiltonian $H_P$
(\ref{HPnew})).  The existence of the extended eigenstate of
$\tilde{H}(\tau)$ whose properties do not depend on a particular
instance of NPP follows directly from Eq.~(\ref{v}) that
involves only a self-averaging quantity
$\chi(\Omega',\Omega,\lambda)$. This quantity varies smoothly over
the broad range of partition residues $|\Omega'|,|\Omega| \alt
\langle E\rangle$ and does not allow for the compression of the
wave-packet $v(\Omega_{z})$ on the much smaller scale ${\cal
O}(2^{-n})$. This gives rise to an eigenstate with probability
amplitude of individual states $|{\bf z}\rangle$ that depends
smoothly on energy in this range.

\begin{figure}[t]
\includegraphics[bb=50 20 341 320,width=3.2in ]{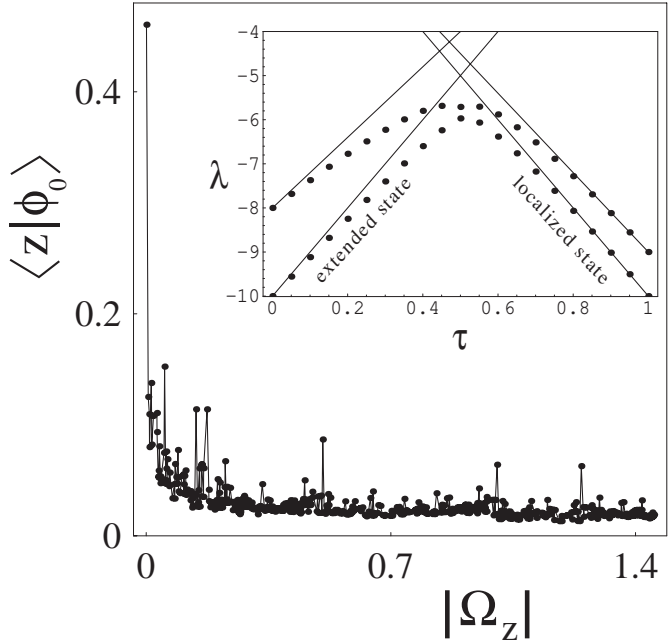}
\caption{\label{fig:2level} Dots correspond to the plot of the
ground state amplitude $\langle {\bf z}|\phi_{0}\rangle$
{\it vs} partition residue $|\Omega_{\bf z}|$ evaluated at the avoided
crossing point $\tau=\tau^*$ (thin lines connecting the dots are for
display purposes). Simulations are done for the
randomly sampled instance of NPP with  $n=10$ and $b=20$;
the corresponding value of $\tau^*\approx 0.5$. In
simulations we relax the condition (\ref{criterion_ld})  and the
value of  $M$ in (\ref{omegaK}) is set automatically to be an integer closest to
$\log_{2}\sum_{j=0}^{n}a_j$ (cf. (\ref{omegaK})).  Insert:
Dotted curves are the plots of the two lowest eigenvalues of
$H(\tau)$ {\em vs} $\tau$  for the same instance of NPP as in the
main figure.  Solid lines that start at $\tau=0$ correspond to
$\lambda=(1-\tau)n+k$ with $k=0,1$ (cf. (\ref{li})). Solid lines that ends at
$\tau=1$ correspond to $\lambda=\tau\,\varepsilon_{k}$ with
$k=0,1$ (cf. (\ref{lf})). }
\end{figure}
\noindent
\subsubsection{Analysis of the general case}
The above picture of avoided-crossing remains qualitatively the
same when the condition (\ref{criterion_ld}) is relaxed (cf.
insert in the Fig.~\ref{fig:2level}). Away from the
avoided-crossing point, $\tau < \tau^*$, the  ground state
wavefunction $v(\Omega_{\bf z})$ and energy
$\lambda_{0}^{i}(\tau)$ are obtained directly from Eq.~(\ref{v})
with replacement  $\Phi \approx \overline{\Phi}$ and
Eq.~(\ref{chi}) taken into account. Because the spectral function
$\chi(\Omega,\Omega',\lambda)$ changes only slightly on the scale
$E_{\min}={\cal O}(n^{1/2}2^{-n})$ the wave packet $\sum_{\bf
z}v(\Omega_{\bf z} )|{\bf z}\rangle$ remains extended,
$|v(\Omega_{\bf z}|={\cal O}(2^{-n/2})$,  and therefore
$\Phi_{0}={\cal O}(2^{-n/2})$.

Beyond the avoided-crossing point, $\tau >\tau^*$, the ground
state is localized near ${\bf w}$ and eigenvalue branch
$\lambda_{0}^{f}(\tau)$ is obtained from Eq.~(\ref{Phi0}) with
r.h.s. set to zero (cf. Sec.~\ref{sec:ld}). The point
$\tau=\tau^*$ is located at the intersection of the two branches
$\lambda_{0}^{i}(\tau)\approx\lambda_{0}^{f}(\tau)$ and the level
repulsion  is of the order of the overlap factor between the
extended and localized states
\begin{equation}
g_{\min}\sim \sum_{{\bf w}\in {\cal L}_0}v(\Omega_{\bf w})\sim
2^{-n/2}.
\end{equation}
Ground-state wavefunction $\phi_{\bf z}$ at the avoided-crossing
is shown in Fig.~\ref{fig:2level} for modest value of $n$, but the
separation into slowly- and rapidly-varying parts (\ref{ansatz})
is clearly seen.

\begin{figure}[t]
\includegraphics[width=3.3in]{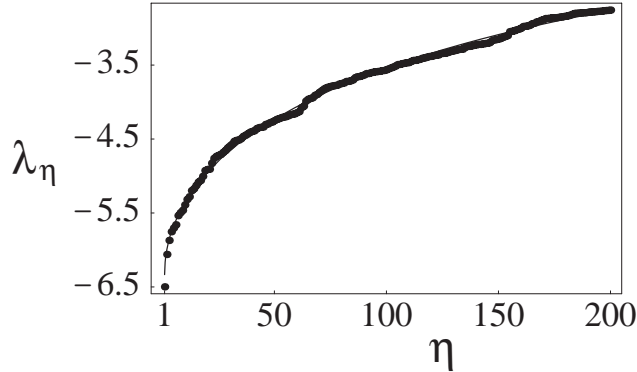}
\caption{\label{fig:energies} Dotted line is plot of
$\lambda_{\eta}$ {\em vs} $\eta$ at the avoided-crossing
$\tau=\tau^{*}$. It is obtained from the numerical solution of the
stationary Schr{\" o}dinger  equation for the same instance of NPP
as in Fig.~\ref{fig:2level}. Solid line is a square-root fit
$\lambda=-6.3+0.35\,\eta^{1/2}$ (solid line is almost
undistinguishable from the dotted line).}
\end{figure}

We did not perform a direct  numerical study of the dependence of
$g_{\min}$ on $n$ since we only simulated adiabatic eigenvalues
for small instances of NPP. We argue, however, that even for a
fixed $n$ the scaling of $g_{\min}$ with $n$ can be inferred from
the shape of the cumulative density of states
\begin{equation}
\eta(\lambda)=\int_{0}^{\lambda} dx\,
\sum_{k=0}^{k_m}\delta\left(\lambda_{k}-x\right ),\qquad
k_m=2^n-1, \label{cum}
\end{equation}
\noindent where $\lambda_{k}\equiv \lambda_{k}(\tau)$ are
eigenvalues of $H(\tau)$ (\ref{adiab}). These eigenvalues are
plotted in Fig.~\ref{fig:energies} near the avoided-crossing
$\tau=\tau^*$ where the spectrum of $\lambda_{k}$ is
quasi-continuous. The shape of the plot is well approximated by
the square-root function:
\begin{equation}
\lambda_{\eta} \approx {\rm
const}+\left(\frac{\eta}{\eta_m}\right)^{1/2},\qquad \eta_m={\cal
O}(2^n). \label{sqr}
\end{equation}
\noindent It is clear that for $\eta \approx  1$ we have
$\lambda_{\eta}\sim 2^{-n/2}$ which corresponds to
Eq.~(\ref{gmin}). Note that this qualitative analysis is based on
the assumption that  the asymptotic properties of $\lambda_{0}$
for large $n$ can be inferred from the behavior of
$\lambda_{\eta}$ for $\eta\gg 1$.

\section{\label{sec:udo} Simulations of time-dependent  Schr{\" o}dinger equation}

We also study the complexity of the algorithm by numerical
integration of the time-dependent Schr{\" o}dinger equation with
 Hamiltonian $H(t)$ and initial state $|\psi(0)\rangle$ defined
in Eqs.~(\ref{H}),(\ref{V}),(\ref{omegaK}),(\ref{HPnew}). Here we
relax the condition $\omega_M \ll \langle E\rangle$ used above in
the analytical treatment of the problem; in simulations the value of
$M$ is set automatically to be an integer closest to
$\log_{2}\sum_{j=0}^{n}a_j$ (cf. (\ref{omegaK})). We introduce a
complexity metric for the algorithm, $C(T) = (1+T)d_0/p_{0}(T)$
where $p_{0}(t)=\sum_{{\bf w}\in {\cal L}_0}|\psi_{\bf w}(t)|^2$.
A typical plot of $C(T)$ for an instance of the problem with
$n$=15 numbers is shown in the insert of Fig. 4. At very small $T$
the wavefunction is close to the symmetric initial state and the
complexity is $\sim 2^{n}$. The extremely sharp decrease in $C(T)$
with $T$ is due to the buildup of the population $p_0(T)$ in the
ground level, ${\cal E}_{\bf z}=\varepsilon_{0}$, as quantum
evolution approaches the adiabatic limit. At certain $T=T_{\rm min}$
the function $C(T)$ goes through the minimum: for $T
> T_{\rm min}$ the decrease in the number of trials $d_0/p_0(T)$ does
not compensate anymore for the overall increase in the runtime $T$
for each trial. For a given problem instance the \lq\lq minimum"
complexity $C_{\rm min}=C(T_{\rm min})$  is obtained via one
dimensional minimization over $T$. The plot of the complexity
$C_{\rm min}$ for different values of $n$ in Fig. 1 appears to
indicate the  exponential scaling law, $C_{\rm min} \sim 2^{0.8n}$
for not too small values of $n \agt$ 11.
\begin{figure}[th]
\includegraphics[width=3.3in]{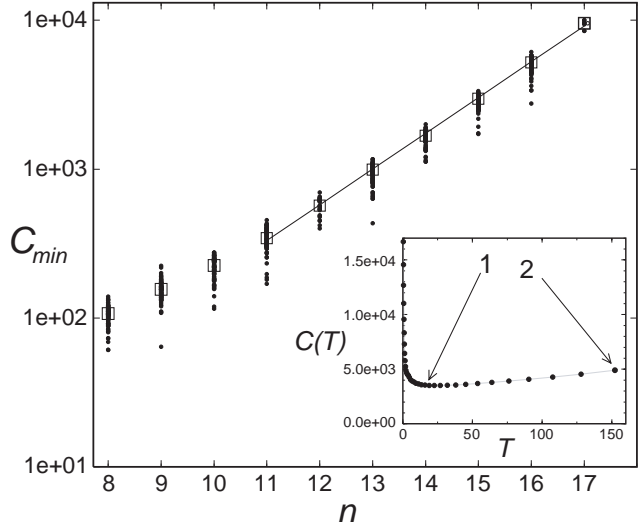}
\caption{\label{fig:complexity} Logarithmic plot of $C_{\rm min}$
{\em vs} $n$ for randomly generated instances of NPP with 25-bit
precision numbers. Vertical sets of points indicate results of
different trials ($\sim$ 100 trials for each $n$, except $n$=17
with 10 trials). Median values of $C_{\rm min}$ are  shown with
rectangles. Linear fit to the logarithmic plot of median values
for $n$ between 11 and 17 is shown by the line and gives $\ln
C_{\rm min} \approx$ 0.55n ($C_{\rm min} \sim 2^{0.8 n}$) . Very
close result is obtained for the linear fit if all data points are
used instead of the median values.  Insert: plot of $C(T)$
{\em vs} $T$ for n=15, precision b=25 bits, $d_0$=22. Point $1$
indicated with the arrow refers to the minimum value of complexity
at $T=T_{\rm min}=22.67$ where the total population of a ground
level $p_0(T_{\rm min})=0.15$. Point $2$ refers to the value of
$T$ where $p_0(T)=0.7$.}
\end{figure}

\section{Discussion}
In conclusion, we have developed a general method for the analysis
of avoided-crossing phenomenon in quantum spin-glass problems and
used it to study the performance of the quantum adiabatic
evolution algorithm on random instances of the Number Partitioning
problem.  This algorithm is viewed as a \lq\lq quantum local
search'' with matrix elements of the Green function $G_r$
($r=1,\ldots,n-1$) giving the quantum amplitudes of the
transitions with different number of spin flips $r$. Our approach
is similar to the analysis of a quantum diffusion in a disordered
medium with the  model of disorder  defined by the one- and
two-point distribution functions $P(\Omega), \, P_{r,{\bf z}}(
\Omega')$.

 We have shown that the conditional distribution of
partition
  residues $P_r(\Omega'|\Omega)$ in the neighborhood of a given string
  formed by all possible $r$-bit flips depends on the value of the
  partition residue for that string but {\it not} on the string itself.
  This is a specific property of the random Number Partitioning problem.

We used the above property to describe a quantum diffusion in the
energy space (Eq.~(\ref{v})). This reduction in the dimensionality
leads to the formation of the eigenstate  which is extended in the
energy space. Near the avoided-crossing the adiabatic ground state
changes from extended to mostly localized near the solution to the
optimization problem. Because the extended and localized state
  amplitudes are nearly orthogonal to each other the repulsion between
  the corresponding branches of eigenvalues (the minimum gap) is
  exponentially small, $g_{\rm min} \sim n\,2^{-n/2}$, and the run
  time of the algorithm scales exponentially with $n$.  Analytical
  results are in qualitative agreement with numerical simulations of the
  time-dependent Schr{\" o}dinger equation for small-to-moderate instances of the
  Number Partitioning problem ($n\leq 17$).

One can show  that the effect of quantum diffusion in
reduced-dimensional
 space that leads to the formation of the extended state can also occur in other random NP-complete
  problems \cite{SST}.
 The method developed in this paper will  be applied to
 study the performance of
 continuous-time quantum algorithms for different random
 combinatorial optimization problems.
Also the present framework can be applied to the analysis of
quantum annealing algorithms for combinatorial optimization
problems \cite{kadowaki,santoro}. This is a classical algorithm
that is conceptually very close to the quantum adiabatic evolution
algorithm considered above \cite{kadowaki1}. The former uses the
Quantum Monte Carlo method to simulate on classical computers a
partition function and ground-state energy of a quantum system
with slowly varying Hamiltonian that merges at the final moment
with the problem Hamiltonian of a given classical optimization
problem.
 Among other possible
  applications of our method is the analysis of tunneling phenomenon in the
  low-temperature dynamics of random magnets.

We note that  the specific property of the Number Partitioning problem
(that distinguishes it from the other NP-complete problems) is a very weak
    dependence of $P_r(\Omega'|\Omega)$ on $\Omega$ for not too large
    values of $\Omega', \Omega\ll \sqrt{r(n-r)}$ that takes place  for
    all values of $r\in [1,n-1]$. This rapid fall-off of correlations during the
    local search (both classical and quantum) is a reason that the exponential complexity of optimization algorithms for the Number Partitioning problem
    can be seen already  for the relatively small values of $n\lesssim 15$
    (cf. Fig.\ref{fig:complexity}).

    Finally, our analysis of sub-harmonic resonances in the Fourier
    transform $I(s)$ of the distribution function $P(\Omega)$ suggests  a
    possible connection between NPP and the integer factorization problem.
    If, for a given set of $a_j$'s, there is a number $q$ that
    satisfies the condition (\ref{gcd2}) then dividing all numbers $a_{j}$
    by $q$ we obtain a new instance of NPP with numbers $k_j=a_j/q$
    that will be completely equivalent to the old one. It is important that the
    precision of the numbers $k_j$ is restricted by $b-\log_{2}q$. If
    the value of $q$ is sufficiently large, $\log_{2}q\gg b-n$, then
    $k_j$'s correspond to a low precision instance of NPP, i.e. to the
    computationally easy phase mentioned in Sec.~\ref{sec:NPP}.
    This is exactly the case when sub-harmonic resonances become
    substantial.  One can fix the parameter $\xi=b/n\gg 1$ in a
    high-precision (computationally hard) case and compute, for randomly
    generated instances $\{a_j\}$  an {\it approximate greatest
      common divider}, i.e.  a largest number $q$ that satisfies
    (\ref{gcd2}). The distribution of these numbers determines a fraction
    of high-precision instances of NPP (out of all possible $2^{n\,b}$
    problem instances) that really belong to a low-precision (computationally
    easy) \lq\lq phase''.

Advance knowledge of this information would be of importance
if one is using NPP for encryption purposes \cite{cr}, especially
because NPP is otherwise a very difficult problem for both quantum
and classical computers \cite{Mertens2000}.  It is not obvious at
this stage what the asymptotic  form of this distribution will be in
the limit of large $n$ (cf. Fig. \ref{sec:subharmonics}).

We  are not aware of any  classical algorithm that could
verify if such a number $q$ exists for a given set of $a_j$ in a
time polynomial in both $n$ and $b$. However, on a quantum computer
one can apply a Shor algorithm to test in polynomial time if
strong sub-harmonic resonances exist. This question is deferred to
a future study.

\section{Acknowledgments}
The authors benefited from stimulating discussions with  P.
Cheeseman, R.D. Morris (NASA ARC) and  U. Vazirani (UC Berkley).
 We also acknowledge the help of J. Lohn
(Automated Design of Complex Systems group, NASA ARC) for
providing computer facilities. This research was supported by NASA
Intelligent Systems Revolutionary Computing Algorithms program
(project No: 749-40), and also by  NASA Ames NAS Center.\vfill

\appendix

\section{\label{sec:subharmonics} Sub-harmonic resonances} We note that  function $I(s)$ in
(\ref{I}) can also have additional sharp resonances in the range
$0 < |s| \leq 2^b$. To understand their origin we consider first a
particular case when rational $b$-bit numbers $a_1,a_2,\ldots,
a_n$ all have a   number $q>$ 2$^{-{\rm b}}$ as a \lq\lq common
divisor'', i.e., there exist integers $k_1,\,k_2,\ldots,k_n$
such that
\begin{equation}
\frac{a_1}{k_1}=\frac{a_2}{k_2}=\ldots=\frac{a_n}{k_n}=q.\label{gcd}
\end{equation}
\noindent In this case additional resonances  of $I(s)$ occur at
the multiples of $\pi/q$.
Assume now that  $q$ is no longer  an
exact divisor of numbers $a_j$ but all the residues of the
divisions $a_j/q$ are sufficiently small. Then  contributions from
the additional resonances at  $s \approx m\pi/q$
($m=1,\,2,\ldots$) to the integral in (\ref{I}) can be estimated
as follows (for simplicity we give a result for the case $E\ll
n^{1/2}$):
\begin{eqnarray}
&&P(0)\rightarrow  \frac{2^{n}}{\sqrt{2 \pi n\sigma^{2}(0) }} \,e^{-\gamma(q)}
\,\sum_{m=1}^{\infty}\,\zeta\left(\frac{m\,\pi \eta}{2q}\right)
(-1)^{m p} \label{gcd1} \\  && p=\sum_{j=1}^{n}\left[\frac{a_j}{q}
\right], \qquad \gamma(q)= \frac{\pi^2}{2}\left( \left\{
\frac{a_j}{q} \right\}^2-\frac{a_j}{\sqrt{\pi\,n \sigma^{2}(0)}} \,\left\{ \frac{a_j}{q}\right\}\right) \nonumber
\nonumber
\end{eqnarray}
\noindent Here $[x]$ and $\{x\}$ denote integer and fractional
parts of a number $x$, respectively.  If the  total \lq\lq
dephasing" factor $e^{-\gamma(q)} \sim 1$, then contribution
(\ref{gcd1}) cannot be neglected in the steepest-descent analysis
of  (\ref{I}) (in general, on should keep contributions from all
divisors $q$ with small dephasing factors $e^{-\gamma(q)}$).

\begin{figure}[bht]
\includegraphics[bb=10  10 541 180]{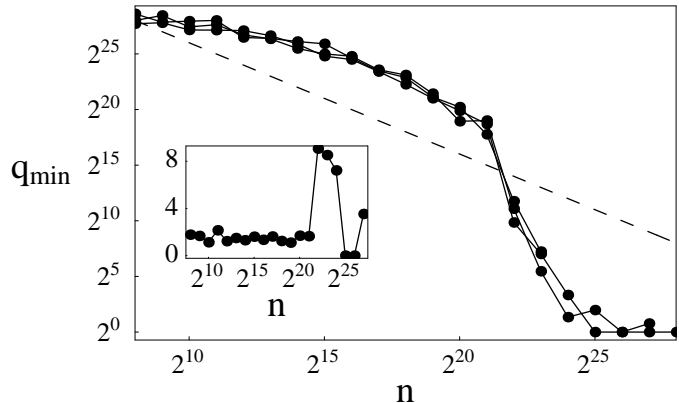}
\caption{\label{fig:gcd} Log-Log plots  of the  mean value of the
largest root
of Eq.~(\ref{gcd2}) $q_{\min}$ {\it vs}  $n$. Three sets of data points
are plotted. Each set of points represents averaging over
25 randomly generated instances of NPP. Precision of the random numbers $a_j$
is 30 bits and the value of $\gamma_c=0.5$. Dashed line corresponds to the plot of $const\times 2^{-n}$ {\it vs} $n$.  Insert: Variance of the $\log_{2}q$   {\it vs } $n$ based on 25 sample points for each $n$.
Distribution of $q_{\min}$ values become very broad when the mean 
drops to $q_{\min}\sim 1$.
}\end{figure}

We note that  the  window function $\zeta\left(\frac{m\,\pi \eta}{2q}
\right)\sim 1$ for $q\gg 2^{-n}$ and it decays to zero at smaller
values of $q$. 
We studied numerically the greatest root $q_{\max}$ of the the algebraic 
equation 
\begin{equation}
 \gamma(q)=\gamma_c  \label{gcd2}
\end{equation}
\noindent
for a fixed value of $\gamma_c \alt 1$. For the sets of random $b$-bit
numbers $a_j$ the dependence of the mean value of $q_{\min}$ on the
problem size $n < b$ is shown in Fig.~\ref{fig:gcd}.  For $n\ll b$ we
have exponential decrease of $q_{\rm max}$ with $n$ and for larger
values of $n\lesssim b$ the value of $q_{\min}$ steeply drops to 1.
According to the discussion above,  in order to neglect the saddle-points
with $s>0$ in (\ref{I}) (additional resonances) the value of $q_{\min}$ should satisfy
the following
condition in the asymptotic limit $b\rightarrow \infty$:
\begin{equation}
q_{\rm max} \lesssim
\max \left[2^{-n},2^{-b}\right], \qquad  1\ll n\ll b,\label{qmin_cond}
\end{equation}
\noindent
with  $\gamma_c$ fixed at some small constant value.
Because  the precision $b$ that we used in the simulations was not very high
(limited by machine precision) it is not possible  
to obtain   the asymptotic form of the dependence 
of $q_{\min}$ on $n$ in the range given in (\ref{qmin_cond}). 
Neither we can describe the shape of the plot in Fig.~\ref{fig:gcd}
analytically in that range. However, it appears from the figure
that  the condition (\ref{qmin_cond}) is 
satisfied for sufficiently large $n$. 

\section{\label{sec:conddist}Properties of the conditional  distribution of signed residues in NPP}

We  perform the summation over the  spin configurations in
Eq.~(\ref{rhoOmegaOmega}) with Eq.~(\ref{rhoOmegazr}) taken into
account. Similar to the derivation of Eq.~(\ref{I}) we use
integral representation for delta function  and obtain
\begin{equation}
P_r(\Omega,
\Omega')=\binom{n}{r}^{-1}\,\int_{-\infty}^{\infty}\int_{-\infty}^{\infty}\frac{ds\,ds'}{4\pi^2}
\,\zeta\left(\frac{\Delta\Omega s}{2}\right)\,
\zeta\left(\frac{\Delta\Omega' s'}{2}\right) \,\,\sum_{{\bf J}}
 U_{\bf J} (s,s'),\label{U}
\end{equation}
\noindent \vspace{-0.2in}
\begin{equation}
U_{\bf J}(s,s')=\prod_{j\in {\bf J}}\cos(a_j(s-s'))\,\times
\,\prod_{j\notin {\bf J}}\cos(a_j(s+s'))\,\times\,e^{ i( s\,\Omega
+ s'\,\Omega' )}.\nonumber
\end{equation}
\noindent Here the  sum is over all possible subsets ${\bf
J}=\{j_1,j_2,\ldots,j_r\}$ of length $r$ obtained from the set
of integers $j=1,2,\ldots,n$. Window function $\zeta(x)$ is defined in
(\ref{I}).   After the change of variables
\begin{equation}
x'=s+s',\qquad x=s-s',\label{x}
\end{equation}
\noindent we obtain from (\ref{U}) that $U_{\bf J}(s,s')$
factorizes into a product of two terms
\[ U_{\bf J}(s,s')\,=\,{\cal V}_{\bf J}(x)\,\widetilde{\cal V}_{\bf J}(x')\]
\noindent
\begin{equation} {\cal V}_{\bf J}(x)=\exp\left(\frac{i
x(\Omega-\Omega')}{2}\right)\prod_{j\in {\bf J}}\cos(a_{j}\,x),
\quad \widetilde{\cal V}_{\bf J}(x')=\exp\left(\frac{i
x'(\Omega+\Omega')}{2}\right)\prod_{j\notin {\bf
J}}\cos(a_{j}\,x').\label{VV}
\end{equation}
\noindent In what follows we will analyze several limiting
cases. \\ \underline{$r,n-r \gg 1$}:\\
\nopagebreak
In this case both functions
${\cal V}_{\bf J}(x)$ and $\widetilde{\cal V}_{\bf J}(x')$ are
very steep and similar to the analysis in Sec.\ref{spp.den}
integrals in (\ref{U}) can be evaluated by the steepest descent
method. With the appropriate choice of the coarse-graining windows
$\Delta\Omega,\,\Delta\Omega'$ in (\ref{U}) (see below)
contribution to the integrals comes from the vicinity of the point
($x=0, x'=0$). Near this point we use
\begin{equation}
\prod_{j\in {\bf J}}\cos(a_{j}\,x)\approx \exp\left(-\frac{r
(x\,\sigma_{\bf J})^2}{2}\right),\qquad \prod_{j\notin {\bf
J}}\cos(a_{j}\,x)\approx \exp\left(-\frac{(n-r)
(x'\,\overline{\sigma}_{\bf J})^{2}}{2}\right)
\end{equation}
\noindent
 where
\begin{equation}
(\sigma_{\bf J})^2 =\frac{1}{r}\sum_{j\in {\bf J}}a_{j}^{2},
\qquad (\overline{\sigma}_{\bf J})^{2} =\frac{1}{n-r}\sum_{j\notin
{\bf J}}a_{j}^{2}.\nonumber
\end{equation}
\noindent Since each sum here contains a large number of terms we
obtain for i.i.d. random numbers $a_1,\ldots,a_n$ (cf.
(\ref{PGauss}))
\begin{equation}
(\sigma_{\bf J})^2\approx \sigma^{2}(0)+{\cal
O}\left(\frac{1}{r}\right), \qquad (\overline{\sigma}_{\bf J})^{2}
\approx \sigma^{2}(0)+{\cal O}\left(\frac{1}{n-r}\right),
\label{sigma}
\end{equation}
\noindent where $\sigma^2=\langle a^2 \rangle$ is given in
(\ref{PGauss}).
 Using Eqs.~(\ref{VV})-(\ref{sigma}) and replacing the window functions in (\ref{U}) by unity,
we compute the Gaussian integrals in (\ref{U}) and  obtain
\begin{equation}
P_r(\Omega,\Omega^{\prime})=\frac{1}{4\pi
\sigma^{2}(0)\sqrt{r(n-r)}} \exp\left[-\frac{1}{8\sigma^{2}(0)}
\left(\frac{(\Omega-\Omega')^2}{r}
+\frac{(\Omega+\Omega')^2}{n-r}\right)\right].\label{rhor}
\end{equation}
\noindent The size of the coarse-graining windows in (\ref{U}) is
chosen to satisfy the conditions
\[2^{-n}\,\binom{n}{r}^{-1}\ll \Delta\Omega\,\Delta \Omega' \ll
\sqrt{r(n-r)}\] \noindent  From Eq.~(\ref{rhor})  and
Eq.~(\ref{PGauss}) one can directly obtain the conditional
distribution function  $P_r(\Omega|\Omega')$
\begin{equation}
P_{r}(\Omega'|\Omega)=\frac{1}{\sqrt{2\pi
n\sigma^{2}(q)}}\exp\left[-\frac{\left(\Omega'-q\,\Omega\right)^{2}}{2n\sigma^{2}(q)}\right].
\label{conditional}
\end{equation}

\noindent \underline{$r=1$; $\quad r=n-1$:}\\ For $r=1$ function
$\overline{{\cal V}}_{\bf J}(x')$ contains a product of $n-1$
terms and is very steep. The corresponding integral over $x'$ in
(\ref{U}) should be taken by the steepest descent method.  However
${\cal V}_{\bf J}(x)$ simply oscillates at  frequencies
$(\Omega-\Omega')/2\pm a_j$ and the integral over $x$ in (\ref{U})
should be evaluated using the corresponding oscillating factors.
In the opposite case $r=n-1$, function ${\cal V}_{\bf J}(x)$ is
very steep and the integral over $x$ in (\ref{U}) should be taken by
steepest descent. But the integral over $x'$ there should be evaluated
using  $\overline{{\cal V}}_{\bf J}(x')$ that oscillates  at the
frequencies, $(\Omega+\Omega')/2\pm a_j$. Finally, one can obtain
using i.i.d. numbers $a_j$'s in $\left[0,1\right]$ interval :
\begin{equation}
P_{r}(\Omega'|\Omega)=\frac{1}{4} \left[\Theta\left
(\Omega\mp\Omega'+2\right)-\Theta\left(\Omega\mp\Omega'-2\right)\right]+{\cal
O}\left(\frac{1}{n}\right),\qquad (r=1,\,n-1).\label{theta}
\end{equation}
\noindent  The minus (plus) sign in (\ref{theta}) corresponds to
$r=1$ ($r=n-1$).  Similarly one can obtain the result
for any fixed value of $r$ or $n-r$ (that does not scale with
$n$). For $|\Omega|,|\Omega'| \lesssim 1$ (\ref{theta}) is reduced
to (\ref{rho1}).

\subsection*{\label{sec:numerics} Numerical simulations of conditional distribution
$P_{r,{\bf z}}(\Omega')$} We compute the following integrated
quantity:
\begin{equation}
Q= \frac{1}{2}\,\int_{0}^{\Omega'}d\eta\, P_{r,{\bf
z}}(\eta),\label{Qe}
\end{equation}
\noindent for  different values of $r$, $\Omega'$ and different
strings ${\bf z}$ with $E_{\bf z} \ll 1$. Numerical results are
compared in the insert to Fig.\ref{fig:density06} with theoretical
result below obtained using $P_r(\Omega'|\Omega)$ from
Eq.~(\ref{conditional})
\begin{equation}
 \frac{1}{2}\int_{0}^{\Omega'}d\eta\, P_{r}(\eta|0) =
{\rm erf}\left(
\frac{\Omega'}{\sigma(q)\,\sqrt{2n}}\right).\label{rhoEEa}
\end{equation}
\noindent Theoretical and numerical curves  nearly coincide
with each other. To accurately compare the normalization factor in
(\ref{conditional}) (see also (\ref{nr}))  we compare the theoretical
results with numerical values of  $P_{r,{\bf z}}(0)$ for different
$r$ and strings ${\bf z}$ corresponding to $E_{\bf z} \ll 1$.  The
results are plotted in Fig.~\ref{fig:density06}.

\end{document}